\documentclass[prl,aps,twocolumn,showpacs]{revtex4}

\pdfoutput=1
\usepackage{graphics}
\usepackage{graphicx}

\begin{document}

\title{The Geometry of Slow Structural Fluctuations in a Supercooled Binary Alloy}

\author{Ulf R. Pedersen}
\affiliation{DNRF centre Glass and Time, IMFUFA, 
Department of Sciences, Roskilde University,
Postbox 260, DK-4000 Roskilde, Denmark}
\affiliation{Department of Chemistry, University of California, 
Berkeley, California 94720-1460, USA}
\author{Thomas B. Schr\o{}der}
\affiliation{DNRF centre Glass and Time, IMFUFA, 
Department of Sciences, Roskilde University,
Postbox 260, DK-4000 Roskilde, Denmark}
\author{Jeppe C. Dyre}
\affiliation{DNRF centre Glass and Time, IMFUFA, 
Department of Sciences, Roskilde University,
Postbox 260, DK-4000 Roskilde, Denmark}
\author{Peter Harrowell}
\email{p.harrowell@chem.usyd.edu.au}
\affiliation{School of Chemistry, University of Sydney, Sydney NSW 2006, Australia}

\pacs{64.70.Rh, 61.20.Ja, 61.43.Fs, 64.70.qd}

\begin{abstract}
The liquid structure of a glass-forming binary alloy is studied using molecular dynamics simulations. The analysis combines common neighbour analysis with the geometrical approach of Frank and Kasper to establish that the supercooled liquid contains extended clusters characterised by the same short range order as the crystal. Fluctuations in these clusters exhibit strong correlations with fluctuations in the inherent structure energy. The steep increase in the heat capacity on cooling is, thus, directly coupled to the growing fluctuations of the Frank-Kasper clusters. The relaxation of particles in the clusters dominates the slow tail of the self-intermediate scattering function.
\end{abstract}

\maketitle

If, on supercooling, a liquid accumulates mechanically stable structures with identifiable geometries, then the explanation of many of the anomalous properties of such liquids, including the glass transition itself, becomes straightforward. The appeal of this possibility has not gone unappreciated. As far back as 1933, Tammann [1] proposed that glasses were the result of the coagulation of some sort of clusters. In the late `70's, Hoare [2] reformulated the Tammann proposal, including an explicit geometrical description of the stable clusters. A number of recent simulation studies [3-7] have, again, revived interest in the possibility of accounting for glassy anomalies in terms of the temperature dependence of specific locally favoured ordering. Two key challenges face any such proposal. The first is to explain how it is that the relevant clusters are stable enough to resist shear yet not capable of translating this stability into growth. Frank's elegant 1952 resolution [8] of this difficulty consisted of pointing out that, in a liquid made up of spherical particles of one size, an icosahedral coordination shell would have a low energy but would be unable to grow extensively without accumulating defects. While a perennially popular proposition [3,6,7,9], the `protection' icosahedral coordination affords against crystallization requires that the structure be rigidly constrained about the Platonic geometry since, as we shall see, distorted icosahedra can be easily incorporated into a crystal.  The second challenge for a cluster approach is to connect the geometrical description provided some choice of clusters with the structures that correspond to low potential energy states -- i.e. to link the geometry with the energetics. In this paper we shall demonstrate an alternate route to stabilizing structural fluctuations against crystal growth, one that does not rely on fortuitous stabilization, and demonstrate the correlation between the structural and the energy fluctuations.

We consider a model alloy consisting of a equimolar binary mixture of Lennard-Jones particles introduced by Wahnstr{\"o}m [10]. The interaction parameters are $\sigma_\textrm{AA}=\sigma_\textrm{AB}/1.1=\sigma_\textrm{BB}/1.2$, $\varepsilon_\textrm{AA}=\varepsilon_\textrm{AB}=\varepsilon_\textrm{BB}$ and $m_\textrm{A}=m_\textrm{B}/2$. Time is given in reduced units of $\sigma_\textrm{AA}(m_\textrm{A}/\varepsilon_\textrm{AA})^{1/2}$ and temperature in reduced units of  $\varepsilon_\textrm{AA}/k_B$.  This binary mixture has been studied in the context of the glass transition by a number of groups [7,10-12]. A time step of $\Delta t=0.005$ was used in the Verlet velocity integrator. Simulations were performed in the constant NVT ensemble using a Nose-Hoover thermostat. The system size was $N_A = N_B = 512$ with a number density of $\rho = 0.75 \sigma_\textrm{AA}^{-3}$. Initial configurations were constructed from a random configuration, cooling from high temperature ($T = 4.16$) to the target temperature in time $t/3$ and then simulated at the target temperature for $2t/3$, where the time $t$ is chosen so that the root mean square displacement of B particles in the latter 2/3 of this run is $5\sigma_\textrm{AA}$. Trajectories for further analysis are run for times $10^3\tau_\alpha$, the structural relaxation time. Crystallizing trajectories were not included in further analysis. Inherent structures were determined via a steepest descent minimization of the potential energy. Particles are considered to be nearest neighbors if they lie within a cut-off distance chosen as position of the first minimum in the appropriate $g(r)$. At $T = 0.624$, this criterion gives us cut-off distances of $r_c^\textrm{AA} = 1.5$, $r_c^\textrm{AB} = 1.6$ and $r_c^\textrm{BB} = 1.7$.

Like all real glass-forming binary alloys, the Wahnstr{\"o}m mixture does, eventually, crystallize. Pedersen et al [13] have reported that this liquid will spontaneously crystallize into a crystal structure with A$_2$B composition. The structure of the crystal is that of MgZn$_2$, which, along with two other A$_2$B structures: MgCu$_2$ and MgNi$_2$, are known as Laves phases and correspond to the densest packings of a binary mixture of hard spheres with a radius ratio of $\sqrt{3/2}=1.2247$ [14]. ($\sigma_\textrm{BB}/\sigma_\textrm{AA}=1.2$ for this model.)  The distinguishing feature of each of the Laves phases is that the minority B component forms a tetrahedral network, either the diamond or londsdaleite (hexagonal diamond) structure. Over 1400 intermetallic compounds occur as Laves phases, making these structures the most common among intermetallics [15]. The unit cell, shown in Fig.\ 1, contains 12 atoms with the smaller A particles (coloured green) coordinated in one of two different (irregular) icosahedral polyhedra, both made up of six A and six B particles. The B particles (coloured grey) all sit in a 16 vertex coordination polyhedral comprised of 12 A particles and four B particles. Note that the composition change associated in crystallizing an AB liquid into an A$_2$B solid are localised in the coordination polyhedra of the larger particles. The MgZn$_2$ crystal consists of a hexagonal diamond network of B particles (similar to arrangement of oxygens in ice Ih) with each BB pair sharing 6 A neighbours.

\begin{figure}
\begin{center}
\includegraphics[width=0.65\columnwidth]{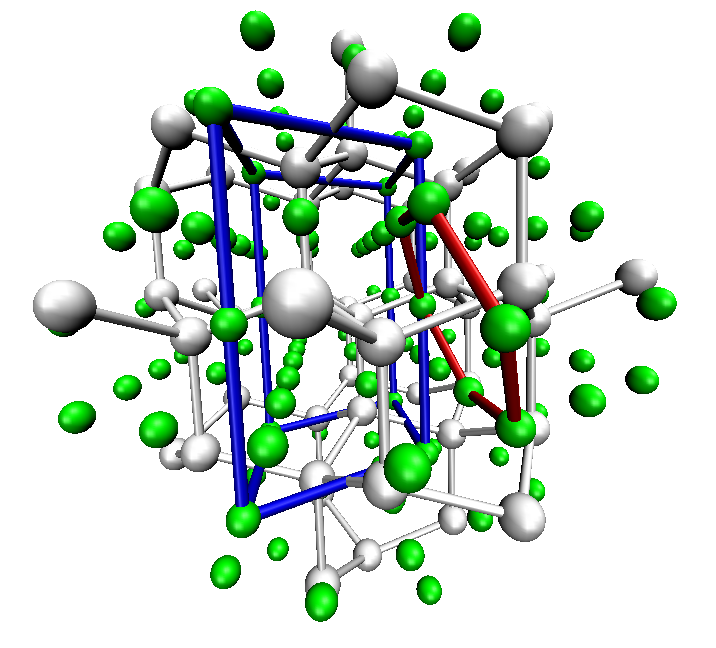} \newline
\includegraphics[width=0.3\columnwidth]{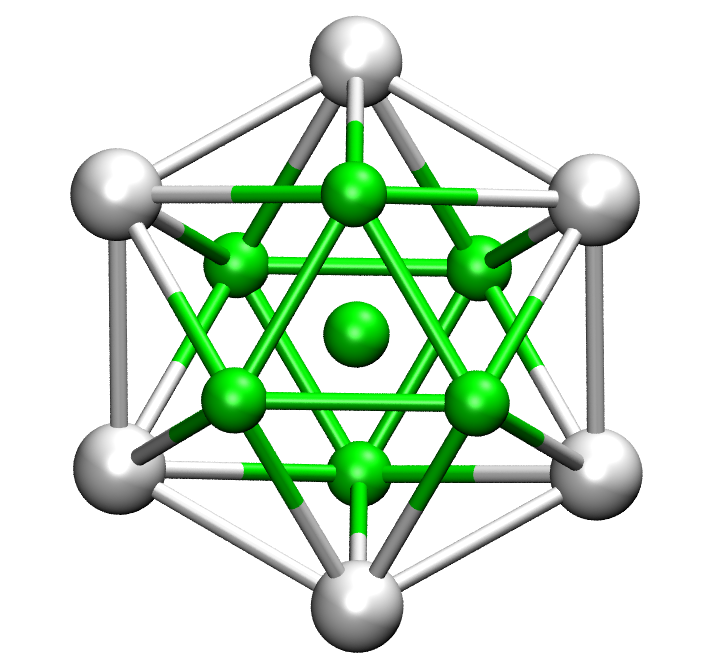}
\includegraphics[width=0.3\columnwidth]{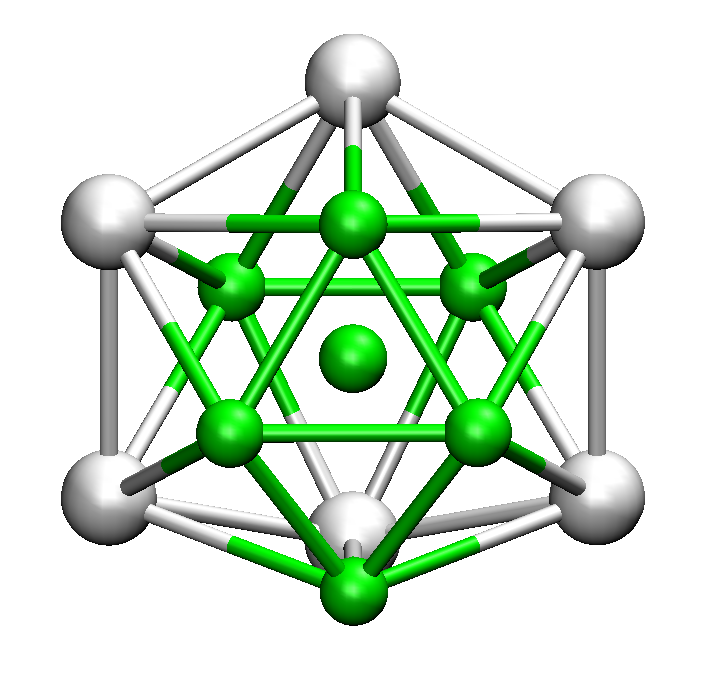}
\includegraphics[width=0.3\columnwidth]{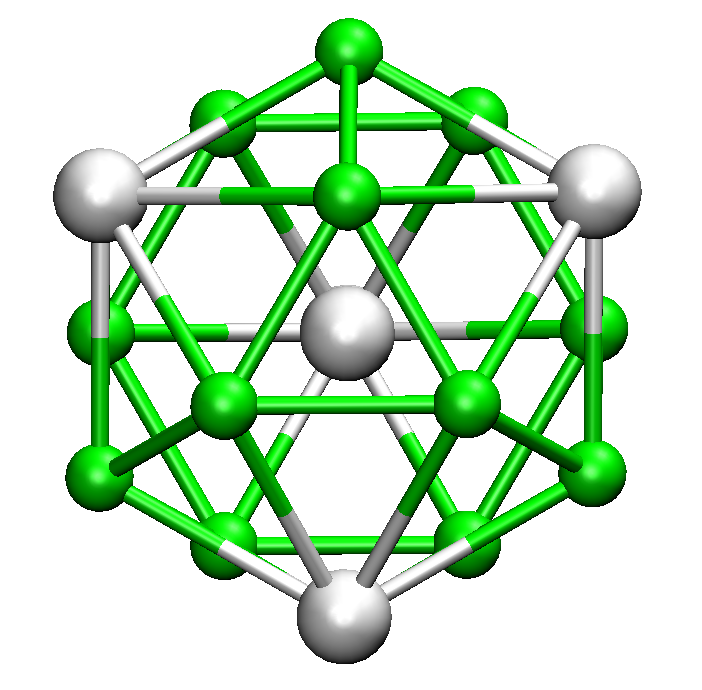}
\end{center}
\caption{Illustration of the unit cell (blue box) of the MgZn$_2$ structure and the three coordination polyhedra found in the crystal. The A (small) particles are green and the B (large) particles are gray (size reduced for visibility). Six A particles of a Frank-Kasper bond are shown connected by red links.}
\end{figure}

Given the complexity of the crystal structure, we direct the reader's attention to the essential feature for the discussion that follows: the unit cell extends beyond the short range (i.e. nearest neighbour) order. This situation is very different, for example, from that of simple liquids that form fcc crystals where the length of short range order and that of the crystal unit cell are one and the same. The size of the unit cell provides one explicit measure of the degree of structural complexity necessary to accommodate particular molecules or atomic mixtures. The significance of the growing difference between short range order and crystallinity (i.e. the appearance of identifiable unit cells) is that such liquids can lower their enthalpy by adopting short range order similar to that found in the crystal without ever having to commit to crystallinity. In the following analysis we present evidence of exactly this phenomenon.

Coslovich and Pastore [6] have established that the Wahnstr{\"o}m mixture contains A-centred icosahedral coordination polyhedra. The fraction of A particles in the centre of such polyhedra grows rapidly on cooling to reach $\sim$25\% at $T = 0.623$. We have confirmed these results and established that the most stable coordination is that comprised of six A and six B particles, just as found in the MgZn$_2$ crystal. To extend the description of the amorphous structure, we shall consider the common neighbour analysis, a method that provides information about both short and intermediate order [7,16]. The elementary objects are nearest neighbour pairs plus those particles that are neighbours to both of particles in the `root' pair. This approach was introduced by Honeycutt and Andersen [17] and provides a finite gallery of geometric objects, bipyramids with the root pair in the axial positions and the common neighbours making up the equatorial sites [18], that are robust in the sense that their size and complexity does not scale as the particle size ratio (unlike the coordination polyhedra).   

In 1958, Frank and Kasper [19] presented a landmark paper on the enumeration of crystal structures built out of triangulated polyhedra. In terms of common neighbour analysis, the four Frank-Kasper polyhedra consist of just 5 and 6 fold bipyramids. The Frank-Kasper polyhedra must have 12 five-fold bonds [19], a requirement that means that the 6-fold `bonds' are a minority component. In ref. [19] the pairs with 6 common neighbours were referred to as `backbone bonds' and were used to build up complex crystal structures by starting with the nets of these backbone bonds. Our strategy is to extend this reasoning to the case of the amorphous alloy and look at building up an extended description of the amorphous states by considering clusters of these backbone bonds. We shall refer to a neighbour pair of B particles with six common A neighbours as a Frank-Kasper (FK) bond. As already mentioned, the MgZn$_2$ crystal consists of tetrahedral network of FK bonds.

Particles are labelled FK particles if they participate in at least one FK bond. The presence of B particles with 4 FK bonds is a necessary condition for crystallinity. In Fig.\ 2 we plot the temperature dependence of i) the average number of B particles in a FK cluster, ii) the average number of B particles in the largest FK cluster, iii) the average number of A particles involved in an FK cluster and iv) the average number of B particles with 4 FK bonds.  The fraction of A particles involved in FK bonds increases sharply on cooling at $T \sim 0.6$, as shown in Fig.\ 2. We note that this temperature is very similar to the value of $T_c$ found on fitting $D \sim (T-T_c)\gamma$ [12] and to the temperature at which an abrupt drop in the average energy of inherent structures explored by the liquid appears [11]. The fraction of B particles with 4 FK bonds, a necessary condition for crystallinity, stays small throughout the studied temperature range. In Fig.\ 3 we show the spatial distribution of FK bonds in a single configuration at $T = 0.599$.

\begin{figure}
\includegraphics[width=1.0\columnwidth]{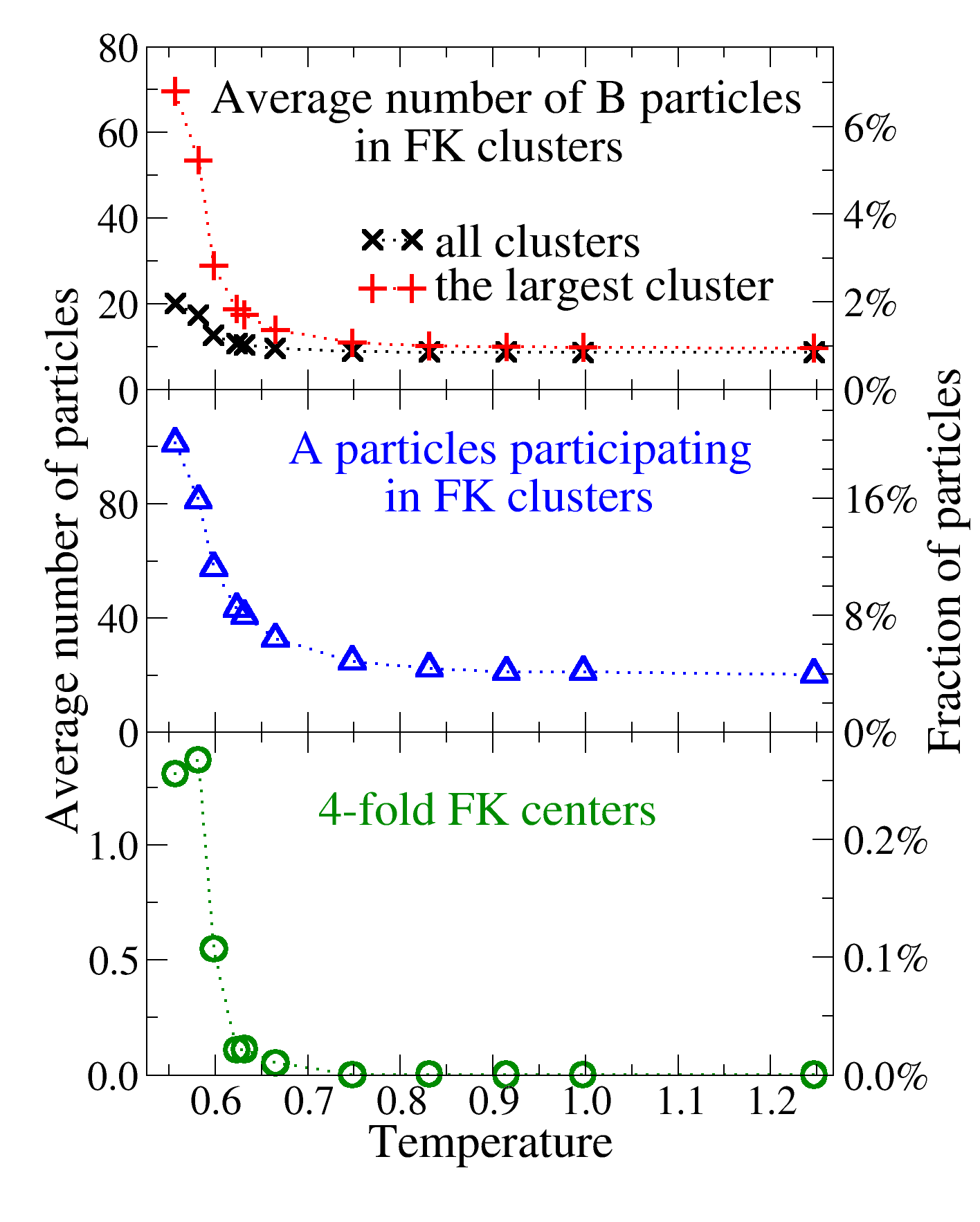}
\caption{ The temperature dependence of the average number of B particles of FK clusters (black XÕs), the average number of B particles of the largest FK cluster (red crosses), the average number of A particles participating in FK bonds (blue triangles) and the average number of B particles with 4 FK bonds (green circles).  }
\end{figure}

\begin{figure}
\begin{center}
\includegraphics[width=0.75\columnwidth]{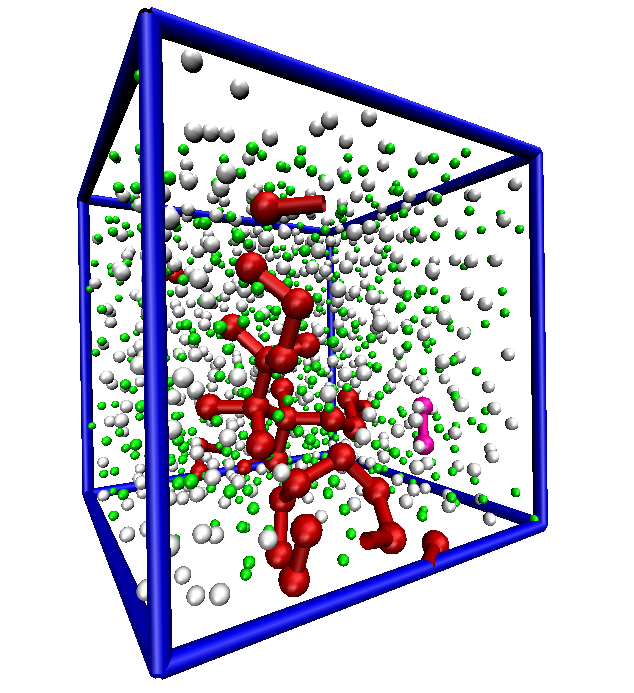}
\end{center}
\caption{A picture of the FK clusters in a supercooled liquid at $T=0.599$. (The inherent structure used corresponds to time $0.5\times10^6$ in Fig.\ 4.) All the FK bonds in a common cluster are depicted as cylinders with the same color. A and B particles are depicted (size reduced for visibility) as green and gray spheres, respectively.}
\end{figure}

We find that the inherent structure energy exhibits persistent dips during trajectories of the metastable liquid. In Fig.\ 4 we show examples of these fluctuations and the associated fluctuations in the size of the largest FK cluster and in the number of B particles with four FK bonds. We find a clear correlation between the transient drop in the inherent structure energy $U_\textrm{is}$ and the appearance of large FK clusters. These large clusters may or may not include a small fraction of 4-fold coordinations. For each instantaneous configuration, the total energy $U = U_\textrm{is} + U_\textrm{vib}$. If we neglect correlations between the fluctuations in $U_\textrm{is}$ and $U_\textrm{vib}$  then we can write the heat capacity as

\begin{equation}
C_{V} \simeq \frac{\langle (\Delta U_\textrm{is})^2\rangle + \langle (\Delta U_\textrm{vib})^2 \rangle}{k_BT^2}
\end{equation}

\begin{figure}
\includegraphics[width=0.85\columnwidth]{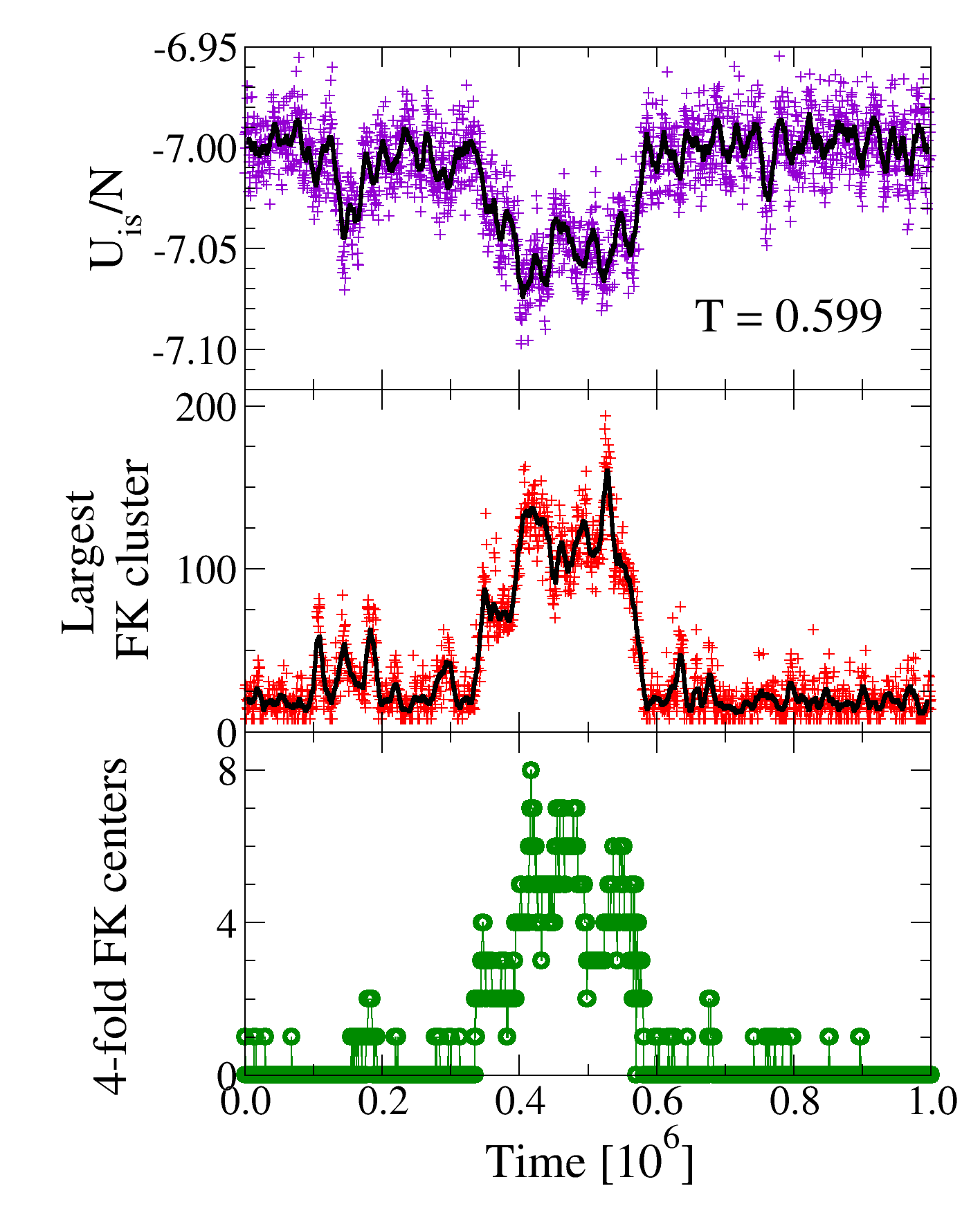}
\caption{The inherent structure energy per particle $U_\textrm{is}/N$, the number of B particles in the largest FK cluster and the number of B particles with 4 FK bonds (collected using the inherent structures) as a function of time during a simulation run. In each case the points correspond to the actual values and, the solid lines are local averages to indicate the trend. Note the clear correlation between the dip in Uis and the increase in size of the largest FK cluster.}
\end{figure}

In Fig.\ 5 we plot $C_V$, $C_\textrm{is}$ (the inherent structure contribution from Eq. 1) and $C_{V}$-$C_\textrm{is}$ as a function of temperature. We find that the rapid increase in $C_V$ around $T \sim 0.6$ is due entirely to the fluctuations in the inherent structure energy which, in turn, we have shown above to be a direct consequence of the fluctuations in size of the FK clusters. The rapid growth of the fraction of particles involved in the FK clusters and the associated energy fluctuations raises the question of the ultimate fate of the supercooled liquid as it accumulates crystal-like short range order. Preliminary results for the A$_2$B liquid, i.e. at the composition as the MgZn$_2$ crystal, show that the liquid at $T = 0.582$ appears to be unstable, its energy decreasing steadily throughout the run [13]. We shall leave a detailed examination of this question for future work.

\begin{figure}
\includegraphics[width=0.9\columnwidth]{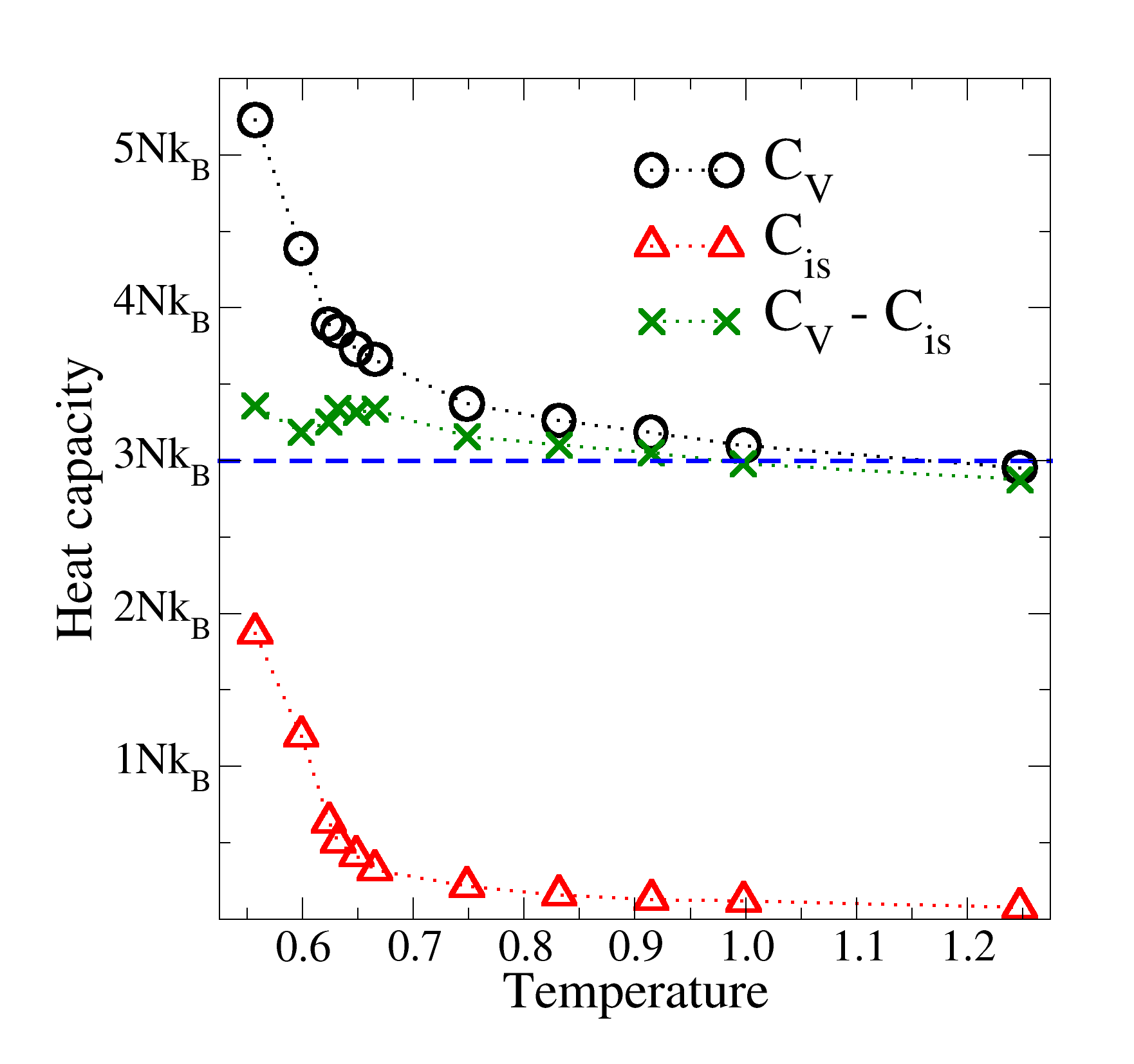}
\caption{ The heat capacity $C_V$ (black circles) versus temperature along with the contribution from fluctuations in the inherent structure $C_\textrm{is}$ (red triangles) and the difference $C_V-C_\textrm{is}$ (green X's). }
\end{figure}

In Fig.\ 6 we have plotted the self-intermediate scattering function from the A particles and the contribution to this relaxation function of those A particles initially in FK clusters. Stretched exponential fits to the curves provide relaxation times for the FK contributions ( 700, 2700, 18000,  in order of decreasing temperature) that are larger than the corresponding times for the non-FK contribution (150, 400, 2300) by a factor that increases from $\sim$4 to 8 on cooling. The stretching exponents are all in the range $0.4$--$0.5$ without any discernable trend. Our geometric criterion successfully identifies the slowest component of the structural relaxation in the supercooled liquid.

We conclude that, in the Wahnstr{\"o}m mixture, the slow relaxation of the structural fluctuations and the appearance of an abrupt increase in the heat capacity on supercooling arise from the appearance of stable clusters whose stability derives from their exploitation of the short range order of the crystal. The clusters are described using the same `backbone' bonds introduced by Frank and Kasper over 50 years ago to enumerate the complex structures of alloy crystals. In the amorphous state, the FK clusters, while `unsaturated' relative to the FK net in the crystal, still achieve a significant decrease in the inherent structure energy. 

These observations constitute a significant advance in the elucidation of the glass transition in this mixture. The kinetic and thermodynamic anomalies that define the glass transition have been associated directly with the development of a specific structure. That the short range order of this structure is the same as that found in the crystal, represents an important break from the tradition of assuming that structure associated with the crystal phase can play no role in the stability of the amorphous phase. The possibility that crystal arrangements can produce an amorphous structure has been discussed previously by Gaskell [20] in the context of metal-metalloid glasses. Tanaka [4] and Procaccia [5] and their coworkers have described models with simple crystal structures in which overtly crystalline domains are associated with the slow down in structural relaxation. In the Wahnstr{\"o}m mixture, we find little sign of the large unit cells that would indicate crystallinity. Configurations of sufficiently low energy to stabilize the amorphous phase are obtained from arranging the short range order found in the crystal into irregular clusters as shown in Fig.\. 3.

\begin{figure}
\includegraphics[width=1.0\columnwidth]{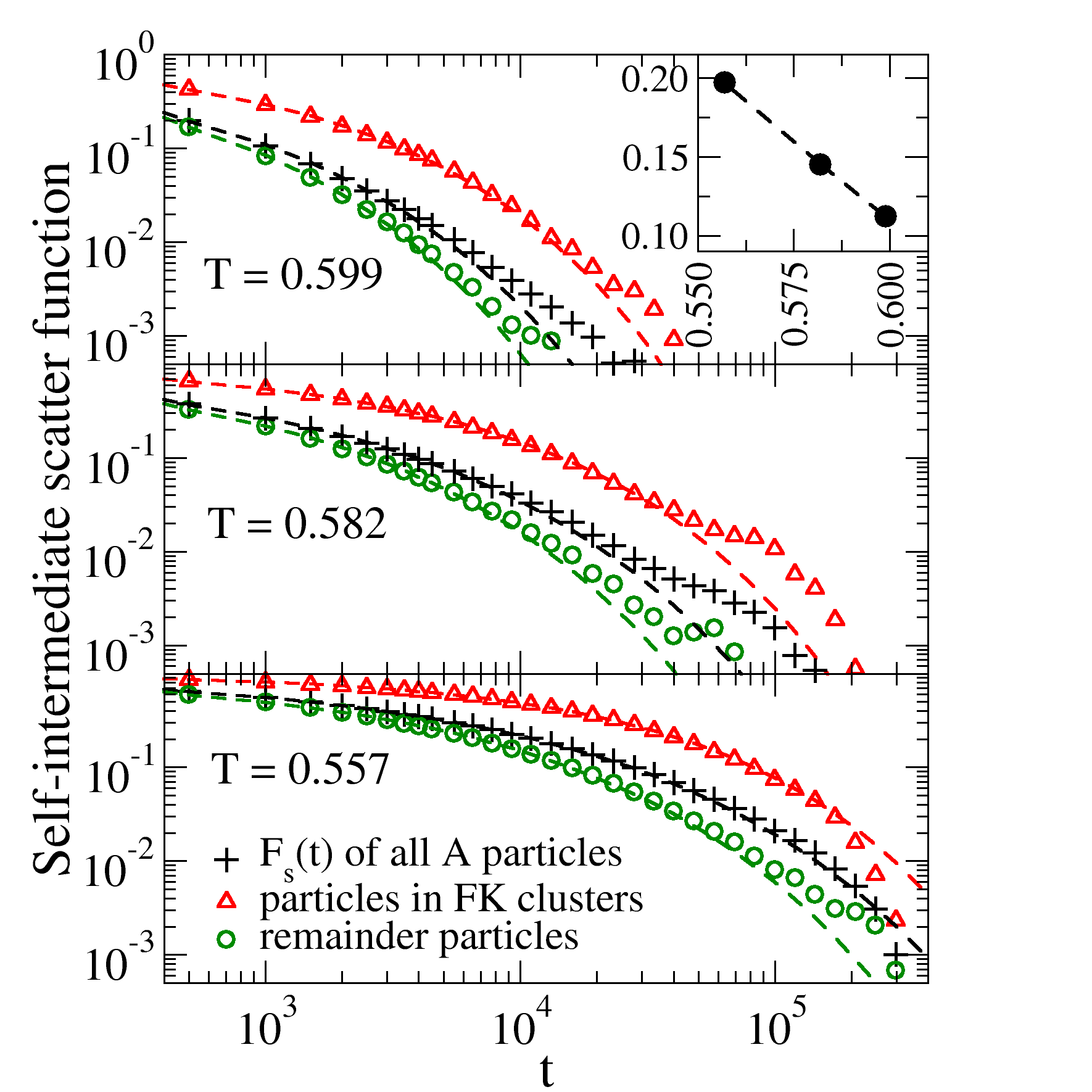}
\caption{Log-log plots of the self-intermediate scattering function $F_s(q=6.47,t)$ from the A particles (black crosses), the contribution $F_\textrm{FK}(t)/F_\textrm{FK}(0)$ due to the A particles initially in an FK cluster (red triangles) and $F_\textrm{n-FK}(t)/F_\textrm{n-FK}(0)$, the contribution due to those A particles not in FK clusters at $t=0$ (green circles) for three temperatures as indicated. Dashed lines represent stretched exponential fits to the data. Insert: The value of $F_\textrm{FK}(t=0)$, the fractional contribution of the A particles initially in a FK cluster to the self-intermediate scattering function, plotted against temperature (the same quantity as plotted in the center panel of Fig.\ 2).}
\end{figure}

In the Wahnstr{\"o}m mixture, frustration of crystallization does not occur in the coordination shell, as proposed by Frank, but in the multiple possible stable arrangements of crystal-like short range order over intermediate lengths. The possibility of this type of frustration arises as a consequence of the fact that the length of the unit cell extends beyond that of the short range order. Crystal nucleation, therefore, requires fluctuations over intermediate length scales. The relationship between crystal nucleation and the pre-existing short range structure is the subject of a report in preparation. How general is the scenario we have described? We don't know. A chemically ordered A$_4$B mixture first studied by Kob and Andersen [21], has been shown [6,16] to have a  short range order in the supercooled liquid dominated by the tricapped triangular prism characteristic of a crystal (the Ni$_3$P structure), but a metastable one -- an observation consistent with the results of this paper. In contrast, preliminary studies of a molecular glass former consisting of a bent triatomic show little correspondence between the short range order of the liquid and the crystal [22]. Clearly we need analysis of more glass forming systems for which the equilibrium crystal structures are known. What we have established in this paper is that the structures that stabilize a glass need not be so different to those that stabilize the crystal, suggesting a possible fundamental continuity between ordered and disordered solid states.

\section{Acknowledgments}

URP and PH gratefully acknowledge many valuable discussions with Toby Hudson on the geometry of crystal lattices. PH would like to acknowledge funding support through the Discovery program of the Australian Research Council. URP, TBS and JCD acknowledge the support of the Danish National Research Foundation's Centre for Viscous Liquid Dynamics Glass and Time. URP is supported by The Danish Council for Independent Research.



\end{document}